# Particle Tracking and Extended Object Imaging by Interferometric Super Resolution Microscopy

Itay Gdor[*a], Seunghwan Yoo[c], Xiaolei Wang[a], Matthew Daddysman[a], Rosemarie Wilton[d], Nicola Ferrier[d], Mark Hereld[b], Oliver (Ollie) Cossairt[c], Aggelos Katsaggelos[c], Norbert F. Scherer[a]

[a] James Franck Institute, University of Chicago, 5801 S Ellis Ave, Chicago, IL, USA 60637;
[b] Mathematics and Computer Science Division, Argonne National Laboratory, 9700 S. Cass Avenue, Lemont, IL USA 60439; [c] Department of Electrical Engineering and Computer Science, Northwestern University, 2145 Sheridan Road, Evanston, IL, 60208 USA. [d] Biosciences Division, A Argonne National Laboratory, 9700 S. Cass Avenue, Lemont, IL USA 60439;

## ABSTRACT

An interferometric fluorescent microscope and a novel theoretic image reconstruction approach were developed and used to obtain super-resolution images of live biological samples and to enable dynamic real time tracking. The tracking utilizes the information stored in the interference pattern of both the illuminating incoherent light and the emitted light. By periodically shifting the interferometer phase and a phase retrieval algorithm we obtain information that allow localization with sub-2 nm axial resolution at 5 Hz.

**Keywords:** super-resolution images, interferometric microscope, biological imaging, particle tracking.

## 1. INTRODUCTION

Optical microscopy is arguably the most common and useful tools to study biological systems, as it is facile straightforward, and (relatively) non-destructive, which allows *in vivo* measurements of living cells and organisms[1]. Regrettably, the resolution of the image acquired by the traditional light microscope is governed by diffraction limit, restricting it to a few hundreds of nanometers in the equatorial plane. This problem is exacerbated in the axial direction, where the effective resolution decreases even more[2].

The spatial resolution of a conventional optical microscope is fundamentally limited by the wavelength of light and the numerical aperture (NA) of the objective lens[1]. While moving to shorter wavelengths does improve the resolution, it is limited by the transparency of the optical element and the biological materials itself, additionally, the short wavelength increase photo-toxicity in live samples[3]. As for improving the NA, in commercially available modern objectives the light collecting angle is close to its practical maximum, so no room is left for further improvement in this direction. Other approaches such as computational deconvolution[4,5], multiphoton microscopy, and confocal [1] techniques have improved the resolution mostly laterally, but also, axially. Nevertheless, axial resolution remains unsatisfactory, especially compared to lateral. The main reason for this inferior axial resolution stems from the fact that a conventional lens collects only a hemispherical wave front moving from the sample toward the lens, and cannot collect angles on the 'back' side of the sample slide. If an imaging system could produce a complete spherical wave front, the focal point, due to symmetry, would be almost spherical as well, resulting in axial resolution that is similar to its lateral counterpart [6]. Thus expanding the detection to collect both front and back emission is equivalent to increasing the aperture angle of the system. Beyond this straightforward improvement, when the optical path lengths of the two (front and back) imaging paths of the microscope are within the coherences length, the two beams of light interfere with each other at the detector. Since phase of the resulted interference pattern is highly sensitive to the position of the emitter, high resolution information in the axial direction is embedded in the interference fringes. Illuminating from both sides increases the contrast by creating an interference pattern in the illumination as well. Encoding high frequency spatial information in the interference fringes is the key principle for widefield $I^2M$/ $I^5M$ [7] and in spot-scanning 4Pi microscopy [8], providing up to sevenfold improvements to axial resolution, and thereby substantially improving the 3D image reconstruction in far-field fluorescence microscopy.

[*] itaygdor@uchicago.edu; phone 773-834-1877; schererlab-sites.uchicago.edu

Our setup, shown in figure 1 is a "in-house" $I^2M/I^5M$ microscope built based on the Gustafsson design.[7] For $I^5M$, the illuminating incoherent LED light was split by a beamsplitter into two equal intensity beams, each beam is sent to one of two opposing objective lenses and focused to the same focal plane where the fluorescent sample is mounted. The sample-emitted florescent light is collected by the same opposing objectives to form magnified images that are combined by the same beamsplitter. The result of the two outputs of the interferometer are imaged and recorded by two EM-CCD cameras. In the $I^2M$ configuration the illuminating incoherent LED light is sent through only one of the objectives (the collection is the same as $I^5M$).

## 2. EXPERIMENTAL

### 2.1 Microscope Experimental Setup

The light source for the illumination in the $I^2M/I^5M$ microscope system was an unpolarized LED (Thorlabs; M490L4), the objectives are a pair of matched $60 \times$ NA 1.3 silicon oil immersion objective lenses (Olympus; UPLSAPO60X). The overlap of the two illumination beams was achieved by controlling the XYZ,θ,φ positions (Mad-city lab; Nano-LP300 nanopositioners) of one of the objectives. The two beam paths matched and scanned by the position of a stage (Newport; XMS100) carrying two mirrors (see Fig. 1). This system maintained phase stability to better than 20 nm (20 nm = 0.2 rad) for 1 hour, this is adequate since each data set takes only 1-3 min to acquire. For 3D imaging the sample was scanned through the image plane using a computer controlled nanopositioner (Mad-city lab; Nano-OP100). Finally, the resulting two outputs of the interferometer were recorded by two EM-CCD cameras (Andor; iXon 888). Data for conventional microscopy for comparisons were acquired by blocking one of the microscope arms. Image data for complete spherical wavefront collection microscopy (CSWCM),[6] were collected using two opposing objectives after changing the optical path of one of the optical arms of the microscope to outside the coherent range of the LED illumination sources so no interference was detected.

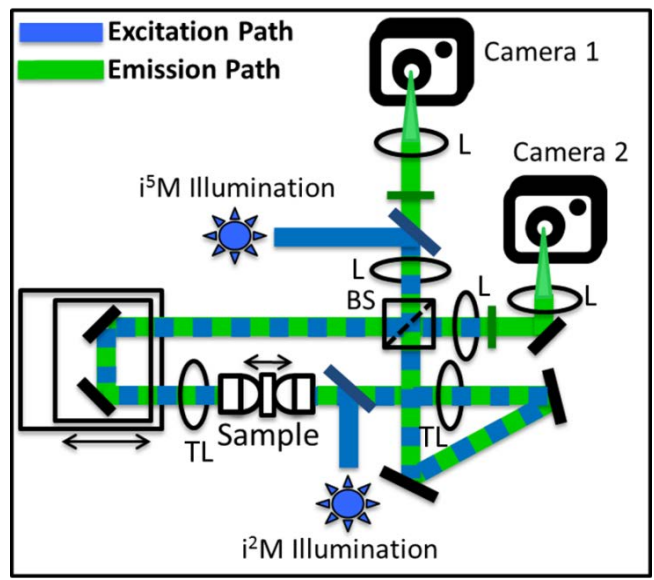

Figure 1. Scheme of $I^2M$ and $I^5M$ setups. In both cases the sample emission (green line) is collected using both objectives and interfered on the cameras, the difference is that in $I^2M$ the sample is illuminated by only one of the objective while in $I^5M$ both objectives are used to create a standing wave in the sample (blue line). TL, tube lens, L, relay and imaging lenses.

### 2.2 Biological Test Samples

Pseudomonas fluorescens SBW25, a Gram-negative bacterium, was isolated from sugar beet leaf in 1989 [9], and is widely studied as a model system of plant colonization and plant-microbe interactions. Its genome has been sequenced and analyzed [10]. SBW25 has plant growth-promoting properties and acts as a biocontrol agent, mediating systemic resistance to bacterial, fungal and oomycete pathogens [11,12]. The SBW25 strain was transformed with a rhizosphere-stable plasmid that drives constitutive expression of the fluorescent protein mNeonGreen [13] into the bacterial periplasmic space. The plasmid is a derivative of pME6031 [14] and was adapted for our use by the introduction of the constitutive promoter Pc, as well as the E. coli TorT signal peptide and a variant of mNeonGreen codon optimized for expression in P. fluorescens. The plasmid is stably maintained in SBW25 during rhizosphere colonization permitting bright labeling of SBW25 in the absence of selective pressure.

To optimize the optical properties of the sample, the bacteria medium was exchanged into an index-of-refraction matched mounting medium composed of Iodixanol solution (SIGMA, D1556) to achieve the desired index (n=1.4). The live cell sample was then placed between to two 100 µm glass coverslips that had been plasma etched with oxygen to charge there surface to allow immobilizing the bacteria.

**2.3 Deconvolution Method**

3D deconvolution of I$^2$M images was achieved by using inverse filtering followed by projection onto the non-negative orthant. Thresholding is used for the inverse filtering to avoid the artifacts from noise, and the thresholds are chosen empirically. The projection onto the non-negative orthant is performed to ensure the non-negativity constraint after the inverse filtering. The point spread function (PSF) was characterized by 200 images of a 170 nm diameter fluorescent bead (Invitrogen Microspheres), acquired at 30 nm focus intervals. The sample was measured with the same focus intervals as the PSF.

**2.4 Pseudo-snapshot interferometric particle tracking.**

Performing super resolution particle tracking; the information embedded in the interferometric measurement of particles moving between the objectives was used to retrieve the phase to localize the particle with nanometric precision. The full protocol is detailed in section 3.3. Briefly, reading the intensity of the light emitted from a particle in two phases enabled us to determine the phase and thus the position of the particle within a single wavelength of the standing wave in the interferometer cavity. This method was demonstrated on a fluorescent bead moving in a known path, resulting in a 3D trajectory map, establishing the reliability of the protocol.

## 3. RESULTS

**3.1 Resolution Characterization**

Figure 2A shows z-axis cut through the raw point spread function (PSF) obtained using a 170 nm fluorescent bead for conventional microscopy, CSWCM, and I$^2$M presented in figure 2 panel B, C and D respectively. The conventional microscope records roughly half the number of photons compared to the other two microscopes as expected as one of the arms that collects the light is blocked, and it provides Poor resolution (the FWHM is 18% wider) when compared to the CSWCM microscope as expected due to the improved NA. The I$^2$M microscopy data exhibit interference patterns that contain the high resolution information (Fig. 2, panel A and D).

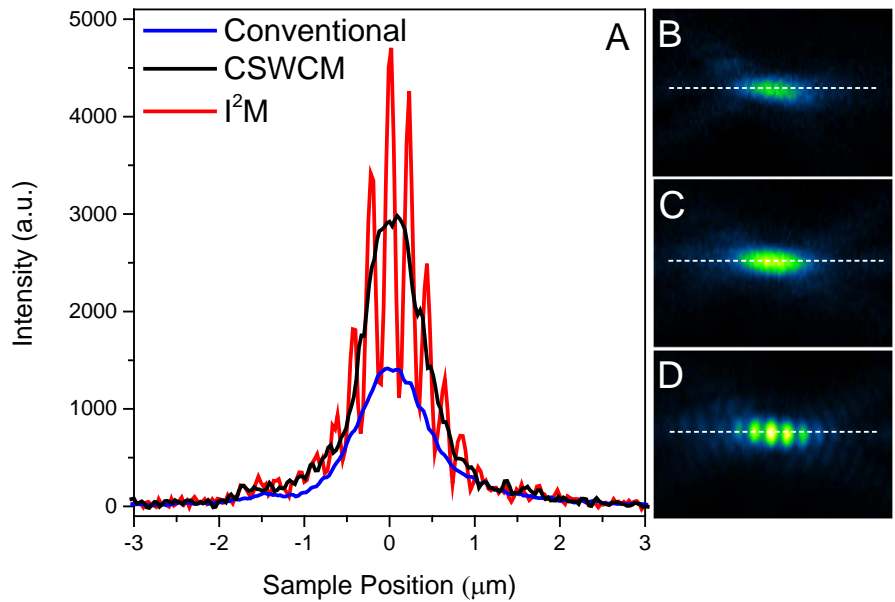

Figure 2. Point spared function for conventional and enhanced resolution florescent microscopy. (A) Z-axis cut and YZ views of a 170-nm diameter fluorescent (B) conventional widefield microscopy, (C) complete spherical wavefront collection microscopy (CSWCM) and (D) I$^2$M.

## 3.2 Live Cell Measurement

We prepared a periplasmically GFP labeled Pseudomonas fluorescens bacteria to demonstrate the abilities of our microscope on live biological systems. Pseudomonas fluorescens are of cylindrical rod shaped cells, 2-5 microns in length and 0.5-1 micron in width [15]. The two sides of the periplasm envelope can be resolved successfully in the XY plane in a conventional light microscope, but not in the XZ and YZ planes, which are too close to the resolution limit of standard microscopy. Therefore this periplasmically labeled system is useful for demonstrating the enhanced resolution in z-plane. All measurements were carried out under the same illumination intensity and detector conditions. The results of I$^2$M measurement before and after deconvolution in comparison to conventional microscopy and CSWCM microscopy are presented in figure 3. The lateral (XY) view of raw data of conventional microscopy, CSWCM and I$^2$M are presented in panels A,E and I respectively, and after deconvolution in panels B,F and J . The axial (YZ) raw data of conventional, CSWCM and I$^2$M are presented in panels C,G and K in respectively and after deconvolution presented in panels D,H and L.

The results show that, aside from the lower signal-to-noise ratio, three different microscopes produce the similar resolution images in the XY plane, before, and especially after deconvolution. This is expected as the enhanced resolution of the CSWCM is not required to resolve the bacteria and the advantage of the I$^2$M will only influence the Z axis. On the YZ plane, however, we can see a clear difference between the three microscopy techniques, as the conventional microscopy, even after deconvolution is unable to resolve the bacteria whereas the I$^2$M shows the expected interference pattern.

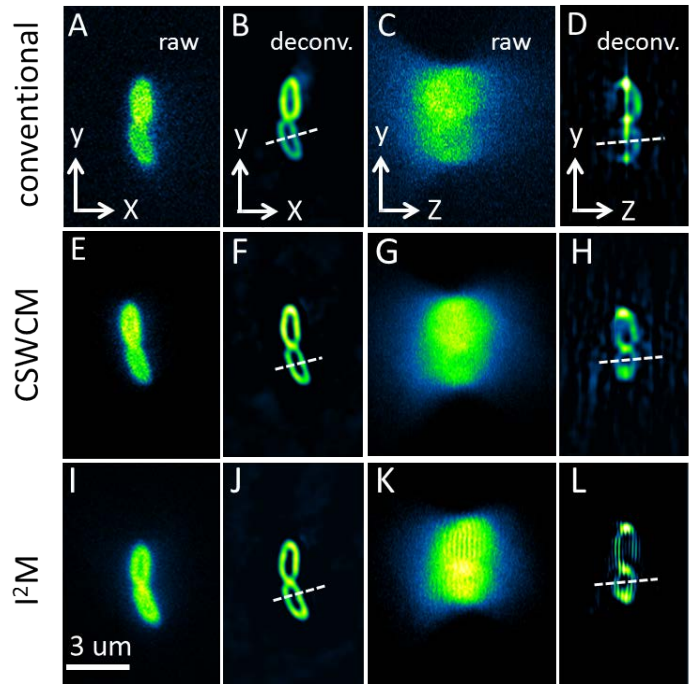

Figure 3. Before and after deconvolution XY and YZ views of Pseudomonas fluorescens labeled with GFP in the periplasm, measured with conventional widefield microscopy (A,B,C,D), CSWCM (E,F,G,H) and I$^2$M (I,J,K,L).

More quantitative information about the imaging and deconvolution result are obtained by examining cross sectional profile through the bacteria profile in the XY plane (figure 4A) and YZ plane (figure 4B). The comparison shows that, as expected, the lateral results before and after deconvolution is not substantially different between the three microscopes as judged by the images in figure 3 and the cuts in panel A of figure 4. In the XY plane all three microscopes were able to resolve the two separate periplasm layers determining their width to be 220±4 nm separated by 550±5 nm of the bacterial cytoplasm. However, the different nature of the I$^2$M is apparent in the axial direction; first, in the raw data, the I$^2$M image exhibits considerably more structure representing the high resolution information (figure 3K).Although a clear separation between the two periplasm layers, which cannot be resolve using the conventional microscope, can be seen. Our deconvolution is not yet ideal and some "striping" is still apparent even after deconvolution. The improved axial resolution in the CSWCM was enough to resolve the two periplasmic layers as seen in figure 3H. The FWHM of the periplasm layer that was extracted using I$^2$M on the axial plane after deconvolution is ca. 350 nm. This value is 60% larger than extracted in the lateral plane and along with the remaining zebra pattern indicate that the relatively simple deconvolution method used here is not sufficient and a better model is required to describe the system more accurately. We are actively working on these issues.

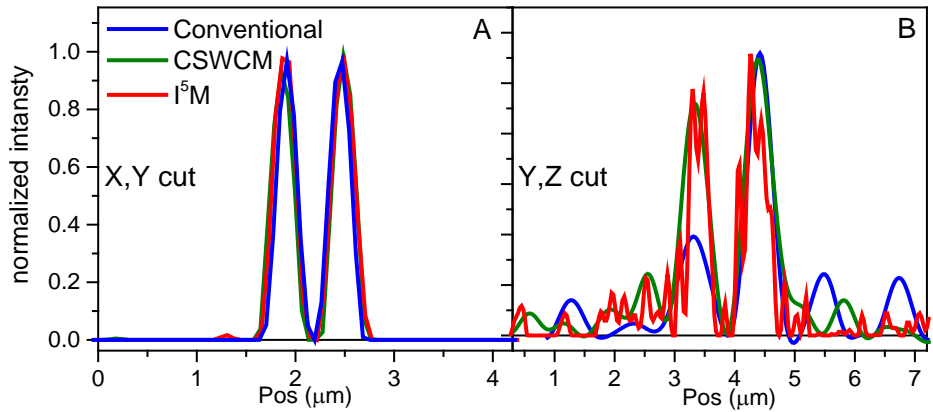

Figure 4. XY (A) and YZ (B) cross sectional profile in the deconvoluted images of Pseudomonas fluorescens presented in figure 3; for conventional widefield microscopy (blue line), CSWCM (black line) and $I^2M$ (red line).

### 3.3 Real-Time 3D tracking using Interferometry Modulation

There are two main limitations in three dimensional tracking that can be addressed using the $I^5M/I^2M$ microscope. One is the need to scan along the Z plane, and the other is the microscope's axial resolving power. By measuring the intensity modulations of light interference in $I^5M$ this resolution can be increased, as constructive or destructive interference depends heavily on the position of the molecule between the two objectives in the two arms of the interferometer. To achieve this high Z resolution, Shtengel et al. [16] recorded three phase-shifted images using a complex custom made 3-way beamsplitter that requires active control. This ingenious optical arraignment allows a "single shot" measurement for particle 3D position determination with sub-20-nm localization precision. However, it also imposes significant constraints on the design and operation of the experimental setup and elaborate optics and precise multi-way beam splitting tailored to each particular dye. Thus, despite the promise of fluorescence self-interference approaches for attaining isotropic 3D resolution, multi-color applications that probe dynamic processes in real-time has been limited. A recent advance in this field comes from Wang et al.[17] who used small modifications of the path length of one of the arms in the microscope to scan over different phases achieving 2 nm localization at 1 Hz.

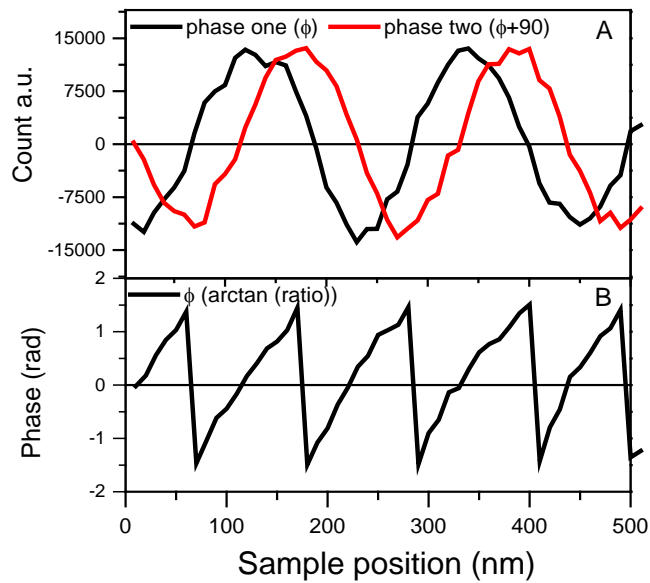

Figure 5. Fluorescence signal and phase reconstruction from our active phase shifting I$^5$M. (A) the difference in emitted light intensity between the two cameras as a function of bead position in two phases (two arms translator position). (B) The measured particle phase as function of bead position extracted using the arctan function.

We found that this localization problem can be simplified further. In practice, achieving interferometric resolution in the Z axis only requires two phase measurements with π/2 shift between them. The mathematical rationalization behind this is that the interference is sinusoidal, thus particle localization within one wavelength simply involved calculating the phase *via* the arctan function. Because sine is not an injective function, measuring the amplitude in one position is not enough, but measuring two phases with π/2 shift between them i.e. measuring sin (φ) and sin (90+φ), is equivalent to measuring sin (φ) and cos(φ), respectively. Dividing the two measurements to obtain a tangent function, which is an injective function, from which φ can be readily extracted. In order to record two measurements with the desired shift between them we used a similar principle as Wang *et al*.[17] and changed the path length of one of the arm in the microscope with the exact amount to cause a π/2 shift in the standing wave. This change in the standing wave was performed repetitively in a square wave fashion throughout the measurement. To demonstrate the method, the integrated intensity emitted from a fluorescent bead was measured as a function of the bead position while moving the bead axially between the objectives. Due to intensity variation along the microscope focus, background noise and readout offset, we found that using a single camera is not sufficient to determine sin (φ) accurately. This issue can be solved by adding a second camera and measuring the second output of the beamsplitter (see figure 1). On first glance the second camera does not add any additional information as a π shift is inherent between both cameras; if camera 1 measures sin (φ) camera 2 will measure sin (π + φ) = - sin (φ). Nevertheless, subtraction of the second camera output, aside from doubling the amplitude (double the number of photons detected), also eliminated of the undesired offset, giving a signal that is symmetric around zero.

The result after separation of the two phases is presented in figure 5A. The use of the two cameras, thus, increases the accuracy of position determination (and reduces the number of variables required to describe the mathematical function). Figure 5B presents the resulting phase extracted by applying the arctangent function to the ratio of the two phase measurements presented in 5A. As can be seen, the phase changes linearly as the particle moves forward. This can be used to generate a calibration curve calculating the relation between the phase and the position, as presented in figure 6. From the slope we extract that translation of 109±1 nm is equal a full π shift, which is expected from the standing wave and the interference pattern in the PSF.

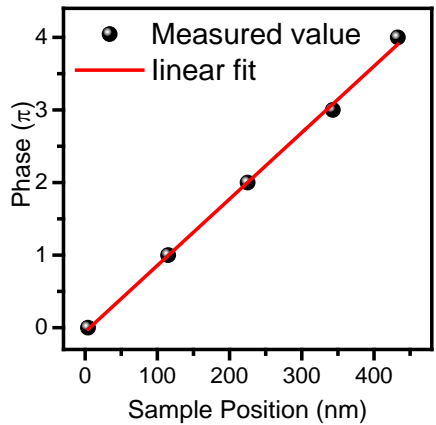

Figure 6. Calibration curve calculating the relation between the phase and the position based on the data presented in figure 5B. The fit gives an error of ±1 nm corresponding to our localization error.

The value calculated from the calibration curve was used to track a fluorescent bead we adhered to a glass coverslip that was moved by a piezo translator in a triangular waveform over 0.5 microns in the axial direction and applied the aforementioned algorithm to extract the bead's axial position. The result compared with the true value given by the translator internal encoder is presented in figure 7A. The results demonstrate that the system can track the motion of the beads with standard error of sub-2-nm in the localization measured at 5 Hz. To demonstrate the ability for 3D tracking, the orientation of the translator that held the sample was slightly altered so the movement in the Z axis was accompanied by movement in the X axis. The resulting tracking is presented in figure 7B. The localization in the XY plane was achieved by 2D Gaussian fitting to find the particle position in every frame. The final bead 3D trajectory is presented in figure 7C. Using the known particle position on the Z plane as a reliable reference allows us to demonstrate our ability to perform three dimensional tracking.

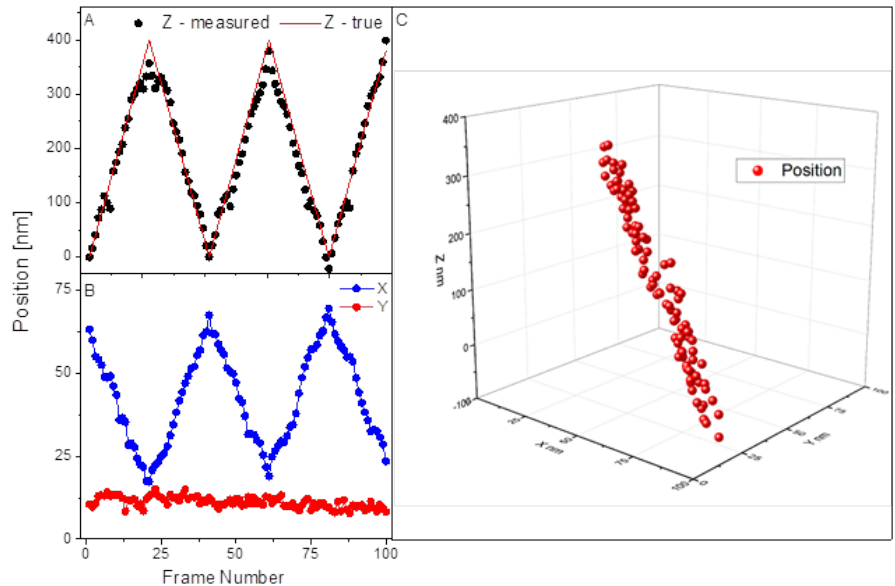

Figure 7. Result of interferometric particle tracking at 5 Hz (A) fluorescent bead moved with a triangle form Z trajectory tracking compared with the true position from the sample translator encoder. (B) XY trajectory of the same bead, extracted using a 2D Gaussian fit. (C) 3D trajectory of the fluorescent bead.

## 4. DISCUSSION

### 4.1 Choosing I$^2$M Techniques for Imaging and Tracking Biological Samples in a Broad Scope

The I$^2$M microscopy described here is similar to other standing wave fluorescence microscopy (SWFM) [24] as it utilizes standing wave illumination in the sample, created by interference of two counterpropagating beams, thus generating interference effects that carry new spatial information. The main advantage of I$^2$M compared to 4Pi confocal microscopy (and its multiphoton derivatives) is that no XY scanning is necessary and a true widefield of the entire XY plane is recorded simultaneously on a CCD camera, allowing more rapid imaging, which is partially important for dynamic biological sample.

In addition, the phase information extracted for a single bead allows high precision localization and enables a relatively simple and reliable tool for particle tracking in 3D as demonstrated in section 3. It is important to stress that this measurements were achieved without the need for any sample scanning in any of the dimensions.

### 4.2 Method Limitations.

Interferometric imaging; While this method provides super resolution images and particle tracking with relatively simple experimental system (in comparison to STED or even 4Pi) it still has several experimental Challenges. 1) Achieving and maintaining alignment of the interferometric signal is time consuming and requires daily realignment. 2) I$^2$M, as do all SWFM microscopes, accesses the focal planes from both sides of the sample, which inherently limits the measurements to thin and transparent samples. In particular, aberration due to refractive index variations within the sample will distort the wave front which diminishes the contrast of the interference pattern. This problem may be aggravated as the sample becomes thicker, we are currently exploring this issue.

Interferometric tracking; This method has several limitations that need to be considered; 1) As the interference is a periodic function, the algorithm always assumes that the particle movement was the shortest distant required to reach the new phase. For example, if the particle moves by $3/2\pi$ between measurements the algorithm will assume that it moved by $-\pi/2$. Thus the system cannot accurately measure particles that move faster than half the wavelength per measurement. 2) the measured length in Z is limited to (a focal area of) approximately 1 micron as illustrated in figure 2, This can be improved by reducing the NA but at the expense of the XY resolution which can be undesirable. 3) The axial diminution of the particle that is being tracked is limited to approximately the emitted light wavelength, as the light emitted from a different part of an elongated object will interfere with different split ratio between the two beamsplitter outputs as they originate from different Z position, thus the contrast of the interference will be lost.

## ACKNOWLEDGEMENTS

This work was supported by funding through the Biological Systems Science Division, Office of Biological and Environmental Research, Office of Science, U.S. Department of Energy, under Contract DE-AC02-06CH11357